\documentstyle[12pt]{article}
\title{MASSIVE CONSTITUENT QUARKS AND UNIFIED DESCRIPTION OF FREEZE-OUT
PARAMETERS:\\ DOES QGP FORM AT LOW ENERGIES?}
\author{O.D. CHERNAVSKAYA,\,\, E.L. FEINBERG,\,\, I.I. ROYZEN} 
\begin{document}
\maketitle 
\centerline{\em{Lebedev Physical Institute of RAS, Leninski prospect 53, 
117333 Moscow,Russia}} 
\centerline{e-mail: $<chernav@lpi.ru>,\quad<feinberg@lpi.ru>,\quad 
<royzen@lpi.ru>$}
\begin{abstract}
The possibility of unified description of hadron multiple production in heavy 
ion collisions over the wide energy interval, from few hundreds MeV/n at 
GSI/SIS through 11 GeV/n at BNL/AGS up to 160 GeV/n at CERN/SPS, has been 
emphasized recently by J. Cleymans and K. Redlich (CR) \cite{CR}. They, and 
somewhat later B. Muller and J. Rafelski \cite{MR}, treated this fact as an 
indication that quark-gluon plasma ($QGP$) can be produced even at very low 
energies of impinging nuclei. In our opinion, it rather witnesses that a 
massive constituent quark (valon) - not massless current quark and gluon! 
- is just what becomes "quite easily" to be unbound, and thus  supports the 
expectation \cite{EF2, Sh} that in course of compression and heating from 
hadronic state to $QGP$ (or, vice versa, of expansion and cooling down from 
$QGP$ to hadronic state) nuclear matter should pass through an intermediate 
phase in between.  
\end{abstract}

\newpage

Analysis of QSI/SIS, BNL/AGS and CERN/SPS data on central collisions of heavy 
nuclei have led CR \cite{CR} to the conclusion that, for any per nucleon 
impinging energy (from the corresponding very wide interval), freeze-out 
points for all species of final hadrons concentrate around a single point in 
the phase space plane $(\mu,T)$, $\mu$ and $T$ being baryonic chemical 
potential and temperature, respectively. If one takes into account the 
observation that final hadrons of each species show up the same mean energy 
about 1 GeV, then these three points fall to the theoretically motivated 
freeze-out curve {it 1} in Fig. 1 which was treated by the authors as a 
boundary between the hadronic and $QGP$ phases (the domains of the plane 
below and above the curve {\it 1}, respectively). 

Of course, the first (hadronic) part of this statement meets no objections, 
whereas the second one deserves a much more careful discussion. Indeed, 
everything, what underlies the remarkable CR observation, is that, at each of 
the above energies, the nuclear matter produced in course of heavy ion 
collision consists of some subhadronic entities mixed in a proper proportion
which coalesce to form hadrons at the Hagedorn \cite{Hag} temperature $T_H$. 
However, 
since this proportion is governed mostly by the differences in masses of the 
entities, irrespectively of whether they are current quarks or valons, it 
is rather insensitive to their mass values themselves. It is why the CR 
results do not necessarily ask for $QGP$ formation - the same unification in 
the $(\mu,T)$-plane is expected to result from coalescence of valons linked 
to current quarks: $Q_u$ and $Q_d$ of the mass $\simeq$ 330 MeV and $Q_s$ of 
the mass $\simeq$ 480 MeV. If the interaction energy is sufficiently high, 
then this pion-valonic state 
\footnote{Pions should present necessarily, although their fraction is rather 
small \cite{ChFR}.} 
comes next to the $QGP$ which is formed at the early stage of interaction, 
whereas, if this energy is rather low, then this state is formed from the 
very beginning of interaction, i.e., no $QGP$ is formed at all. 
Within this pattern of nuclear matter (fireball) evolution, one can, obviously, 
get rid of the questionable (and annoying) assumption of mysterious low 
energy $QGP$. Instead, formation of an chemically and thermally equilibrium 
pion-valonic state ($Q-\pi $) is suggested which precedes the hadronization. 
This would not be even an assumption, if valons were incorporated 
consistently into the QCD formalism. However, all the endeavors to do this 
have been unsuccessful \cite{W}. 

Still one point deserves  mentioning in this connection. After hadronization, 
the final state attractive interaction between hadrons remains still quite 
strong (corresponding $pn$ cross section may be of several hundreds mb), what 
is confirmed by tremendous fraction of final deuterons at low energies, 
$d/p\,\simeq$ 0,37 at SIS. At higher energies, when pions are produced 
numerously, a similar (but, most probably, somewhat lower) effect should be 
caused by the $\rho$-meson resonance. Thus, the general patterns of nuclear 
matter evolution should look like as it is illustrated by curves {\it 2} and 
{\it 3} in Fig. 1., $QGP$ phase being pushed out toward the higher 
temperatures and chemical potentials. 

Actually, the most of what was said above in order to clarify the real 
physical sense of CR observation is nothing else, than the "old news". 
Most probably, E. Shuryak was the first, who summarized a number of 
theoretical indications known at that time and called the attention 
\cite{Sh} to possible existence of a new mass scale of the order of 300 MeV 
in the hadron physics. He expressed also an idea of two separate phase 
transitions - the valonic deconfinement and chiral symmetry restoration. 
Similar concept was considered in \cite{EF2}, where it was noticed that only 
$\simeq$ 3 time increase of density (as compared to the nucleus one) is 
needed to form a configuration of nearly "close packed" (i.e., getting in 
touch) nucleons. 
\footnote{A model calculation \cite{Ch} showed that at $T$ = 0 a
somewhat more dense nucleon closing is needed.}
As the compression goes on, the valons no longer "recognize" their own 
nucleon and thus are getting unbound (more strictly, their confinement radius 
becomes $\simeq\,A^{1/3}$ times larger, $A$ being the number of compressed 
nucleons). Of course, one can never meet such conditions in real nucleus 
collisions
\footnote{However, they could be realized, probably, in neutron star wombs. 
The particular case of $T\,\to$ 0 and very high compression (large $\mu$) 
deserves special discussion, since it is the only one when immediate 
transition between $QGP$ and hadronic phases seems quite probable.}
 - even at SIS, the freeze-out temperature is expected to be about 55 MeV 
and therefore, the hadronization temperature $T_H$ should be still noticeably 
higher. Under such temperatures, nucleons should disintegrate, apparently, 
even at lower compression \footnote{Probably, a mentioning is worthy that 
already long ago the late experimenter from JINR (Dubna) E. Okonov expressed a 
guess \cite{Ok} that observed by him outcome of strange particles was 
suggestive for making the assumption that thermalization and quark 
deconfinement is achieved in nucleus collisions already at laboratory 
energies of few GeV}. The main question of interest is now, what has to 
result from their disintegration:  firstly the quazi-ideal gas of valons 
of a size (0,2-0,4) fm \cite{AQM} (and pions) or immediately the point-like 
current quarks and qluons? For some time, this problem was subjected to a 
quite lively discussion \cite{*d} which then unfoundedly faded. In our 
opinion, the CR observation showed that withdrawal of valons was, at least, a 
premature doing.

In this connection, we would like to remind that presented above qualitative 
picture (see Fig. 1) of phases passed by hot and dense nuclear matter in 
course of its evolution is emerged, in particular, from a certain QCD bag 
model incorporating valons inevitably which was worked out \cite{ChF} in more 
detail. Also a quite good explanation of the observed low (and, maybe, middle) 
mass dilepton yield was obtained \cite{ChFR} from consideration of the 
pion-valonic phase the expanding nuclear matter is assumed to pass through. 

This work is supported by the Russian Foundation for Basic Researches, grants
No.'s 96-15-96798 and 00-02-17250.

~\\
~\\

FIGURE CAPTIONS\\

Figure 1. Chiral transition (solid curve {\it 3}) and hadronization (dashed 
curve {\it 2}) with an intermediate phase in between, preceeding 
the Cleymans-Redlich (CR) freeze-out unified pattern 
(dashed and solid curves {\it 1}). 

\end{document}